\pdfoutput=1
\documentclass[aps,10pt,prl,byrevtex,twocolumn,footinbib,longbibliography]{revtex4-1}
\usepackage{microtype}
\usepackage{graphicx}
\usepackage{amsmath}
\usepackage{amssymb}
\usepackage{color}
\definecolor{myblue}{rgb}{0.153,0.322,0.706}
\usepackage[colorlinks,linkcolor=myblue,urlcolor=myblue,citecolor=myblue,breaklinks=true]{hyperref}

\newcommand{\be}{\begin{equation}}
\newcommand{\ee}{\end{equation}}

\newcommand{\ra}{\rightarrow}
\newcommand{\cL}{\mathcal{L}}
\newcommand{\reals}{\mathbb{R}}
\newcommand{\p}{\partial}
\newcommand{\cH}{\mathcal{H}}
\newcommand{\tI}{\tilde I}
\renewcommand{\SS}{\mathcal{S}}
\newcommand{\T}{\tau}
\newcommand{\hx}{{\hat x}}
\newcommand{\dx}{\dot x}
\newcommand{\z}{\zeta}
\newcommand{\bx}{\bar x}
\newcommand{\bp}{\bar p}
\newcommand{\Di}{\mathcal{D}}

\begin{document}

\title{Anomalous scaling of dynamical large deviations}

\author{Daniel Nickelsen}
\email{danielnickelsen@sun.ac.za}
\affiliation{National Institute for Theoretical Physics (NITheP), Stellenbosch 7600, South Africa}
\affiliation{\mbox{Institute of Theoretical Physics, Department of Physics, University of Stellenbosch, Stellenbosch 7600, South Africa}}

\author{Hugo Touchette}
\email{htouchet@alum.mit.edu, htouchette@sun.ac.za}
\affiliation{National Institute for Theoretical Physics (NITheP), Stellenbosch 7600, South Africa}
\affiliation{\mbox{Institute of Theoretical Physics, Department of Physics, University of Stellenbosch, Stellenbosch 7600, South Africa}}

\date{\today}

\begin{abstract}
The typical values and fluctuations of time-integrated observables of nonequilibrium processes driven in steady states are known to be characterized by large deviation functions, generalizing the entropy and free energy to nonequilibrium systems. The definition of these functions involves a scaling limit, similar to the thermodynamic limit, in which the integration time $\T$ appears linearly, unless the process considered has long-range correlations, in which case $\T$ is generally replaced by $\T^\xi$ with $\xi\neq 1$. Here we show that such an anomalous power-law scaling in time of large deviations can also arise without long-range correlations in Markovian processes as simple as the Langevin equation. We describe the mechanism underlying this scaling using path integrals and discuss its physical consequences for more general processes.
\end{abstract}

\maketitle

%%%%%%%%%%%%%%%%%%%%%%%%%%%%%%%%%%%%%%%%%%%%%%%%%%%%%%%%%%%%%%%%%%%%%

The fluctuations of thermodynamic quantities, such as work, heat or entropy production, are known to play an important role in the physics of molecular motors, computing devices and other small systems that function at the nano to meso scales in the presence of noise \cite{ritort2008,sekimoto2010,jarzynski2011,seifert2012}. The distribution of these quantities is described in many cases by the theory of large deviations \cite{dembo1998} in terms of \emph{large deviation functions}, which play the role of nonequilibrium potentials similar to the free energy and entropy \cite{oono1989,touchette2009,harris2013}. These functions are important as they characterize the response of nonequilibrium processes to external perturbations \cite{baiesi2009,baiesi2009b,maes2011}, general symmetries in their fluctuations known as ``fluctuation relations'' (see \cite{harris2007} for a review), as well as dynamical phase transitions \cite{garrahan2007,chandler2010,hurtado2011b,baek2015,baek2017}.

The definition of large deviation functions involves a limit similar to the thermodynamic limit in which the logarithm of generating functions or probabilities are divided by a scale parameter (e.g., volume,  particle number, noise power, or integration time $\T$) which is taken to diverge \cite{touchette2009}. This applies, for example, to interacting particle systems, such as the exclusion and zero-range processes, which have been actively studied as microscopic models of energy and particle transport \cite{spohn1991,derrida2007,bertini2007,bertini2015b}. In this case, large deviation functions are defined by taking a large-volume or hydrodynamic limit \footnote{The hydrodynamic limit is equivalent, via the macroscopic fluctuation theory \cite{bertini2015b}, to a low-noise limit.}, as well as a limit involving $\T$ when considering time-integrated or \emph{dynamical observables} such as the current or activity \cite{derrida2007,bertini2007,bertini2015b}. 

In this paper, we show that the latter limit must sometimes be replaced by $\T^\xi$ with $\xi\neq 1$ to obtain well-defined large deviation functions. Such an \emph{anomalous scaling} of large deviations arises in many stochastic processes, but it is understood (and now widely assumed) to apply to processes that are non-Markovian or involve constraints that lead to long-range correlations. Examples include random collision gases \cite{gradenigo2013}, disordered and history-dependent random walks \cite{dembo1996b,gantert1998,zeitouni2006,harris2009}, the Wiener sausage \cite{berg2001}, tracer dynamics \cite{krapivsky2014,sadhu2015,imamura2017}, the KPZ equation \cite{doussal2016,sasorov2017,corwin2018}, and branching processes \cite{cox1985b,louidor2015,derrida2017}. Our contribution is to show that the same anomalous scaling can arise without long-range correlations and in processes that are Markovian, ergodic, and non-critical. Moreover, we show that the rate function, one of two important large deviation functions, can be nonconvex, which challenges yet another assumption held in large deviation theory and nonequilibrium statistical physics.

These results apply to a large class of processes, as will be argued, but to illustrate them in the simplest way possible, we consider the dynamics of a Brownian particle described by the overdamped Langevin equation or Ornstein--Uhlenbeck process,
\be 
\label{eqOU}
\dot X_t = -\gamma X_t +\sigma \eta_t,
\ee
where $X_t\in\reals$ is the position of the Brownian particle at time $t$, $\gamma>0$ is the damping, $\eta_t$ is a delta-correlated, Gaussian white noise with zero mean, and $\sigma>0$ is the noise intensity, proportional to the square root of the temperature for a thermal environment. For this process, we consider the dynamical observable to be
\be 
\label{eqconstraint}
A_{\T} = \frac{1}{\T}\int_0^{\T} X_t^\alpha d t,
\ee
where $\alpha$ is an integer assumed to be positive and $\T$ is again the integration time.

Various versions of this model, determined by $\alpha$, have been considered in the context of nonequilibrium systems and turbulence. The case $\alpha=1$, for instance, is related to Brownian particles pulled by laser tweezers, for which $A_\T$ represents the work (per unit time) done by the laser in the harmonic regime \cite{zon2003a}. Alternatively, $X_t$ can be interpreted as the voltage in a circuit perturbed by Nyquist noise, with $A_\T$ then playing the role of dissipated power \cite{zon2004a}. For $\alpha=2$, $A_\T$ is a statistical estimator of the variance of $X_t$, which can be used to measure the damping $\gamma$ or the diffusion constant of Brownian motion ($\gamma=0$) \cite{landais1999,bercu2002,boyer2011,boyer2012}. Finally, the value $\alpha=3$ determines the third moment of $X_t$, related in stochastic models of flow velocity fluctuations to the energy  rate transferred in the turbulent cascade, while higher moments ($\alpha>3$) are important for probing small-scale intermittency \cite{pedrizzetti1994,sreenivasan1997,matsumoto2013,nickelsen2017}. 

We are interested here to study the full probability distribution of $A_\T$ denoted by $P_\T(a)$. In the ``normal'' regime of large deviations, this distribution scales as
\be
P_\T(a)\sim e^{-\T I(a)}
\label{eqldt1}
\ee
for large integration times, $\T\gg 1$, so that the limit
\be
I(a) = \lim_{\T\ra\infty} -\frac{1}{\T}\ln P_\T(a)
\label{eqldtlim1}
\ee
exists and defines a non-trivial function called the \emph{rate function} \cite{dembo1998}. This function is positive and such that $I(a^*)=0$ for the expected value
\be
a^* = \int_{-\infty}^\infty \rho_s(x)\, x^\alpha d x,
\label{eqmean1}
\ee 
obtained from the stationary distribution $\rho_s(x)$ of $X_t$. This implies that fluctuations away from $a^*$ are exponentially unlikely, so that $A_\T\ra a^*$ with probability 1 as $\T\ra\infty$, in accordance with the ergodic theorem. In this limit, $I(a)$ thus characterizes the likelihood of fluctuations of $A_\T$ around $a^*$, in the same way that the entropy characterizes the fluctuations of equilibrium systems around their equilibrium state in the thermodynamic limit (see \cite{touchette2009} for more details on this analogy).

Normal large deviations are found when $\alpha=1$ or $\alpha=2$, and in both cases the rate function is obtained from the dominant eigenvalue of the Feynman--Kac equation for the generating function of $A_\T$. This spectral result is well known \cite{eyink1996a,majumdar2002,fujisaka2007}: it is detailed in \cite{touchette2017} and is briefly summarized in the \hyperref[appSM]{Supplemental Material} (SM) for completeness. The end result is that $I(a)$ is given by a Legendre transform of what is essentially the ground state energy of the quantum harmonic oscillator. From this mapping, one finds a parabolic rate function associated with Gaussian fluctuations of $A_\T$ for $\alpha=1$, and a more complicated rate function describing non-Gaussian fluctuations for $\alpha=2$~\cite{bryc1997}. 

A problem arises, however, when $\alpha>2$. Then the mapping yields a quantum potential which is not confining and, therefore, has no ground state energy for some parameter values. For $\alpha=3$, for example, one finds that the quantum potential is
\be
V_k(x) = \frac{\gamma^2 x^2}{2\sigma^2}-\frac{\gamma}{2}-k x^3,
\ee
where $k$ is the real parameter entering in the generating function of $A_\T$, which is related to the rate function by Legendre transform (see the \hyperref[appSM]{SM}). This potential has no finite ground state energy for any $k\in\reals$ because of the $x^3$ term, which means that the rate function is not related to a ground state energy or dominant eigenvalue. The same applies for any odd integers $\alpha>3$, suggesting that $P_\T(a)$ either does not scale exponentially with $\T$ or that the scaling is exponential but becomes anomalous, in the sense that
\be
P_\T(a)\sim e^{-\T^\xi I(a)}
\label{eqldtgen1}
\ee
with $\xi\neq 1$, and so that $\T$ must be replaced by $\T^\xi$ in the limit \eqref{eqldtlim1} to obtain the correct rate function.

There is no method, as far as we know, that can give the rate function of $A_\T$ in this new scaling regime for arbitrary noise amplitude \footnote{In many of the applications mentioned in the introduction, including the KPZ equation, the large deviation function is obtained in the anomalous scaling via exact representations or mappings based, for example, on Coulomb gases and random matrices. We know of no such mappings for our problem.}. However, we can explore the form of $P_\T(a)$ in the low-noise limit using the well-known saddle-point, instanton or optimal path approximation method, widely used to study noise-activated transition phenomena in equilibrium and nonequilibrium systems \cite{onsager1953,freidlin1984,graham1989,luchinsky1998,nickelsen2011,grafke2015}, including the KPZ equation \cite{kolokolov2008,fogedby2009,meerson2016} and interacting particle systems described in the hydrodynamic limit by stochastic transport equations \cite{derrida2007,bertini2007,bertini2015b}. This approximation is summarized in the \hyperref[appSM]{SM} and leads here to 
\be
P_\T(a)\sim e^{-\SS_\T[\bx]}
\ee
as $\sigma\ra 0$, where $\bx(t)$ is the optimal path or \emph{instanton} that minimizes the action 
\be
\label{eqaction}
\SS_\T[x] = \frac{1}{2\sigma^2} \int_0^\T \big(\dx(t) + \gamma x(t)\big)^2 d t
\ee
of the Ornstein--Uhlenbeck process subject to the constraint $A_\T=a$ in \eqref{eqconstraint}. In our case, $\bx(t)$ is given by the following Euler-Lagrange equation:
\be
\ddot x(t)= \gamma^2x(t) - \beta\sigma^2\alpha x(t)^{\alpha-1}
\label{eqele}
\ee
with free boundary conditions, where $\beta$ is a Lagrange parameter that fixes the constraint $A_\T=a$. Equivalently, we can obtain $\bx(t)$ by solving Hamilton's equations associated with the Hamiltonian,
\be
H(x,p) = \frac{\sigma^2 p^2}{2} - \gamma x p + \beta x^\alpha.
\label{eqH1}
\ee

\begin{figure}[t]
\includegraphics{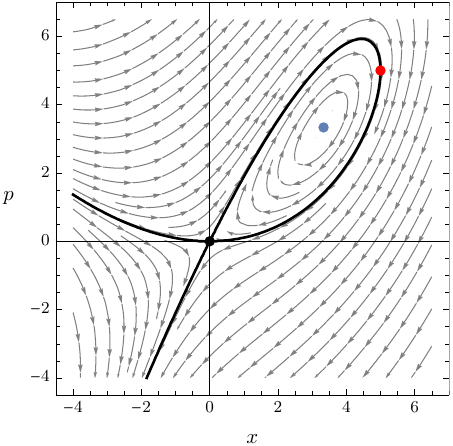}
\caption{Stream vector field of Hamilton's equations describing the instanton in phase space for $\alpha=3$, $\gamma=1$, $\sigma=1$, and $\beta=0.1$. Black line: $H(x,p)=0$ manifold. Black point: Unstable fixed point at the origin. Blue point: Stable fixed point $(x^*,p^*)$. Red point: Turning point $(\hx,p(\hx))$.}
\label{figham}
\end{figure}

We cannot solve these equations exactly for finite $\T$ and $\alpha>2$. However, we find numerically that, as $\T\ra\infty$, $\bx(0)$ and $\bx(\T)$ approach $0$, implying that the associated momentum $p=(\dx + \gamma x)/\sigma^2$ and ``energy'' $H$ also vanish. The infinite-time instanton thus evolves in phase space on the $H=0$ manifold, as shown in Fig.~\ref{figham}: it escapes the unstable origin, performs a loop on the positive part of the zero-energy manifold in finite time, before returning to $(0,0)$. As a result, we can express the action as
\be
\SS_\T[\bx] = \oint_{H=0} p\, dx + \beta \T a.
\label{eqrateham}
\ee
The line integral can be calculated exactly and so can the Lagrange parameter as a function of the constraint $A_\T=a$ (see the \hyperref[appSM]{SM}). Combining these, we find that $\SS_\T[\bx]$ is proportional to $\T^{2/\alpha}$, so that $P_\T(a)$ has the form \eqref{eqldtgen1} with $\xi=2/\alpha$, and
\be
\label{eqIscal}
I(a) = \frac{\SS_\T[\bx]}{\T^\xi} = c\,\frac{\gamma^\frac{\alpha+2}{\alpha}}{\sigma^2} a^\frac{2}{\alpha},
\ee
where
\begin{equation}
c = \pi^\frac{\alpha-2}{2\alpha}
\left[\textstyle\frac{2}{\alpha+2}\frac{\Gamma\big(\frac{2}{\alpha-2}\big)}{\Gamma\big(\frac{\alpha+2}{2\alpha-4}\big)} +\frac{1}{\alpha-2} \frac{\Gamma\big(\frac{\alpha}{\alpha-2}\big)}{\Gamma\big(\frac{3\alpha-2}{2\alpha-4}\big)}\right]
\left[\textstyle\frac{\alpha-2}{2}\frac{\Gamma\big(\frac{3\alpha-2}{2\alpha-4}\big)}{\Gamma\big(\frac{\alpha}{\alpha-2}\big)}\right]^\frac{2}{\alpha} 
\label{eqpref1}
\end{equation}
is a constant prefactor. In particular,
\be
I(a) = \left(\frac{9}{10}\right)^\frac{1}{3} \frac{\gamma^\frac{5}{3}}{\sigma^2} a^\frac{2}{3}\quad\text{and}\quad I(a) = \left(\frac{4}{3}\right)^\frac{1}{2} \frac{\gamma^\frac{3}{2}}{\sigma^2} a^\frac{1}{2}
\ee
for $\alpha =3$ and $\alpha=4$, respectively. Note that, for simplicity, we only give the result for $a\geq 0$, since $A_\T\geq 0$ when $\alpha$ is even, whereas $I(-a)=I(a)$ when $\alpha$ is odd due to the symmetry of the process.

\begin{figure}[t]
\includegraphics{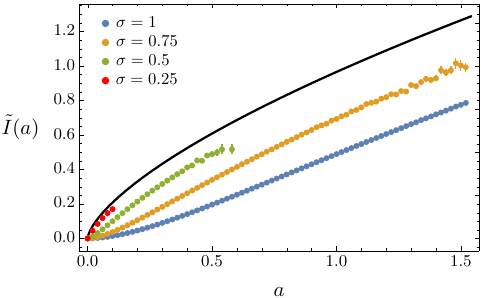}
\caption{Scaled rate function $\tilde I(a)=\sigma^2 I(a)$ for $\gamma=1$ and $\alpha=3$, plotted for $a\geq 0$. Black curve: Low-noise result \eqref{eqIscal}, which is independent of $\sigma$ after rescaling. Data points: Monte Carlo results for different noise amplitudes. Error bars are shown on all points but are in most cases too small to be seen (see text and the \hyperref[appSM]{SM}).}
\label{figrtf1}
\end{figure}

This exact expression for the rate function is our main result. Although it is valid in the limit $\sigma\ra 0$, we show in Fig.~\ref{figrtf1} that it gives a good approximation of the ``true'' rate function obtained by Monte Carlo simulations for $\sigma>0$, up to around $\sigma=0.5$. To obtain this plot, we simulated $10^9$ paths of the Ornstein--Uhlenbeck process using the Euler--Maruyama scheme, and transformed the histogram of $A_\T$ for different $\T$ according to the large deviation limit \eqref{eqldtlim1} with $\T$ replaced by $\T^\xi$, so as to get an estimate of $I(a)$ (see the \hyperref[appSM]{SM}). We also plot $\tI(a)=\sigma^2 I(a)$ rather than $I(a)$, since the low-noise prediction \eqref{eqIscal} is independent of $\sigma$ under this rescaling.

The results are found to converge for $\T\gtrsim 20$ or $\T\gtrsim 30$, depending on the noise amplitude considered, and confirm that $P_\T(a)$ scales anomalously according to \eqref{eqldtgen1} with the predicted $\xi=2/\alpha$. There are very few data points for $\sigma=0.25$, since we are dealing with rare fluctuations that are suppressed exponentially in $T$ and $1/\sigma^2$, but those obtained confirm the function obtained in \eqref{eqIscal}, which is, interestingly, non-convex and homogeneous (or scale-free). The $\T$ scaling with $\xi=2/\alpha$ is consistent with the fact that there is no mapping to the quantum problem, since it implies that the generating function of $A_\T$ diverges for all $k\neq 0$. This can also be seen by noting that, since $\xi<1$ for $\alpha>2$, we get $I(a)=0$ if we use the ``wrong'' limit shown in \eqref{eqldtlim1}. The Legendre transform of that zero rate function diverges for all non-zero values of the conjugate parameter $k$, which is what the quantum problem predicts in the absence of bound states (see the \hyperref[appSM]{SM}).

This applies to any odd $\alpha>2$, for which the mean $a^*$, as given by \eqref{eqmean1}, vanishes since $\rho_s(x)$ is even in $x$. For even values of $\alpha>2$, the situation is slightly more involved. Then $A_T\geq 0$ and, for $0\leq a<a^*$, $A_T$ has normal large deviations with $\xi=1$ \cite{fatalov2006,fatalov2009}, since the quantum problem has a bound state, from which we can obtain the exact rate function, as described in the \hyperref[appSM]{SM}. For $a>a^*$, however, we have anomalous large deviations with $\xi=2/\alpha$ and a rate function given, in the low noise limit, by our general result \eqref{eqIscal}, which predicts that the mean is $0$, consistently with the fact that $a^*\ra 0$ as $\sigma\ra 0$.

\begin{figure}[t]
\includegraphics{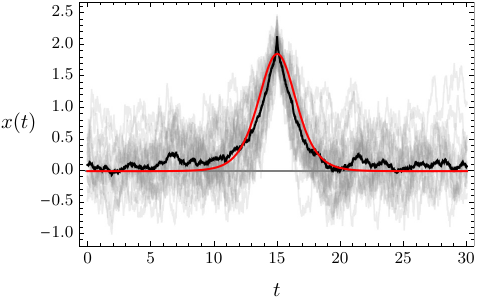}
\caption{Typical paths of the process (in gray) satisfying the constraint $A_T=a$ found by direct Monte Carlo simulations, compared with the instanton (in red) computed numerically. Parameters: $\gamma=1$, $\sigma=0.5$, $\T=30$, and $a=0.45\pm 0.02$. The maximum of each instanton is translated to $t=15$. Black curve: Average instanton.}
\label{figinst1}
\end{figure}

To illustrate the physical meaning of the instanton, we show in Fig.~\ref{figinst1} typical paths of the process with $\gamma=1$ and $\sigma=0.5$ leading to a given fluctuation $A_\T = a$ after $\T=30$, the observed convergence time. For these parameters, we found 28 out of $10^9$ simulated paths reaching the value $A_\T = 0.45\pm 0.02$, which lies on the green curve in Fig.~\ref{figrtf1}. Since fluctuations  can happen in simulations anywhere in the whole time interval $[0,\T]$, we compare these paths by translating their maximum at the time $\T/2$ where the instanton has its own maximum. This also allows us to compute an average fluctuation path which can be compared with the predicted instanton \cite{grafke2015}. 

All the paths are in good agreement, as can be seen, which shows that the low-noise theory correctly predicts how fluctuations are created dynamically by escaping to a position $\hx$, which scales like $(a\T)^{1/\alpha}$, over a finite time proportional to $1/\gamma$. It can be verified (see the \hyperref[appSM]{SM}) that approximating this escape path from $x=0$ by two exponentials with rate $\gamma$ reproduces the correct $\T$ scaling of the action, though not the exact, low-noise expression of the rate function. Similar results are obtained for other values of $A_\T$ and $\alpha>2$, provided that $\T$ is large enough and $\sigma$ is small enough.

The instanton that we find is similar to those arising in the Kramers escape problem \cite{onsager1953}, underlying many noise-induced transition phenomena \cite{gardiner1985}. The essential difference is that we consider a ``global'' constraint $A_\T=a$ rather than a ``local'' constraint for the escape that a process reach a given point or set in time. The instanton is also related to condensation phase transitions in interacting particle systems, such as the zero-range process, in which an extensive number of particles accumulate on a spatial site \cite{grosskinsky2003,harris2013b,chleboun2014}. Here, we find ``temporal condensates'' in the form of trajectories for the fluctuations of $A_\T$ that are localized in time compared to $\T$ and whose height scales with $\T$. A related condensation was reported recently in the context of sums of random variables, which can be dominated in some cases by a single, extensive or ``giant'' value \cite{szavits2014,szavits2014b,zannetti2014,corberi2015,corberi2017,szavits2015}. 

The results that we have presented show that temporal condensation phenomena can arise in simple continuous-time processes, and are not necessarily associated with power-law distributions, as found in \cite{zannetti2014,corberi2015,corberi2017}. They also show, more remarkably, that anomalous large deviations can arise without long-range correlations, non-Markovian dynamics or disorder, and can be linked generally to a breakdown of the quantum formalism used to calculate rate functions. As such, they are expected to arise in other reversible systems for which this formalism can be applied whenever the quantum potential related to the process and observable \cite{touchette2017} does not have a finite ground state.

The problem remains to find the exact rate function of $A_\T$ in the anomalous regime for arbitrary noise amplitudes. Most analytical methods rely on the normal scaling of large deviations and, as a result, cannot be applied. This includes the quantum mapping, as mentioned, but also the so-called contraction principle \cite{dembo1998}. There is a possibility that one can obtain $I(a)$ by finding the exact generating function of $A_\T$ via, e.g., a time-dependent Feynman--Kac equation \cite{boyer2011} in which $k$ is scaled with time. However, if $I(a)$ is non-convex, then even this method will not work, since the Legendre connection between generating functions and rate functions is lost \cite{touchette2009}.

The same limitations apply to numerical methods developed recently to compute rate functions efficiently. Except for the direct Monte Carlo method used here, all methods, including cloning \cite{giardina2006,lecomte2007a,giardina2011} and importance sampling \cite{bucklew2004,touchette2011,chetrite2015}, work by reweighting trajectories exponentially with time in a normal way. In this sense, the model proposed here should serve as an ideal toy model to develop new analytical and numerical methods that are applicable to physical systems with anomalous large deviations, including the many non-Markovian and disordered processes mentioned in the introduction.

%%%%%%%%%%%%%%%%%%%%%%%%%%%%%%%%%%%%%%%%%%%%%%%%%%%%%%%%%%%%%%%%%%%%%
\begin{acknowledgments}
We are grateful to S.\ Sanbhapandit, S.\ Majumdar, R.\ Chetrite, J.\ Meibohm, and A.\ Krajenbrink for useful discussions, and also thank M.\ Kastner for computer access. Support was received from NITheP (Postdoctoral Fellowship) and the National Research Foundation of South Africa (Grant no.~96199). Additional computations were performed using Stellenbosch University's HPC1.
\end{acknowledgments}

%%%%%%%%%%%%%%%%%%%%%%%%%%%%%%%%%%%%%%%%%%%%%%%%%%%%%%%%%%%%%%%%%%%%%
\bibliography{masterbib}

%%%%%%%%%%%%%%%%%%%%%%%%%%%%%%%%%%%%%%%%%%%%%%%%%%%%%%%%%%%%%%%%%%%%%
\appendix
\section*{Supplemental material}
\label{appSM}

\subsection*{Large deviations of dynamical observables}

The most common approach used to obtain the rate function of observables of Markov processes proceeds from the G\"artner--Ellis Theorem \cite{dembo1998} by calculating the limit function
\be
\lambda(k)=\lim_{\T\ra\infty} \frac{1}{\T}\ln \langle e^{\T kA_\T}\rangle,\quad k\in\reals,
\ee
referred to as the \emph{scaled cumulant generating function} (SCGF). The $\langle\cdot\rangle$ denotes the expectation over the trajectories of the process. Following this theorem, if $\lambda(k)$ exists and is differentiable, then $P_\T(a)$ has the scaling shown in \eqref{eqldt1} and the rate function is given by the Legendre--Fenchel transform of $\lambda(k)$:
\be
I(a) = \sup_{k\in\reals} \{ka -\lambda(k)\}.
\ee
In many cases, this transform reduces to the more common Legendre transform; see \cite{touchette2009}.

To obtain the SCGF, we note that the generating function
\be
G(x,\T) = \langle e^{\T k A_\T}\rangle_x
\ee
calculated from all trajectories started at $X_0=x$ evolves according to the partial differential equation
\be
\label{eqfk1}
\p_\T G(x,\T) = \cL_k G(x,\T),
\ee
which is a version of the Feynman--Kac formula involving the linear operator $\cL_k$, called the tilted generator \cite{touchette2017}. For the process \eqref{eqOU} and observable \eqref{eqconstraint}, this operator has the form
\be
\cL_k = -\gamma x\p_x +\frac{\sigma^2}{2}\p_{xx} +k x^\alpha.
\ee

At this point, we obtain $\lambda(k)$ by expanding the evolution of $G(x,\T)$ in the eigenbasis of $\cL_k$. Under appropriate conditions on the spectrum of $\cL_k$ (see \cite{touchette2017}), this evolution is dominated exponentially by the largest eigenvalue $\z(\cL_k)$ of $\cL_k$, i.e., 
\be
G(x,\T) \sim e^{\T \z(\cL_k)},
\label{eqgf1}
\ee
so that $\lambda(k)=\z(\cL_k)$.

The operator $\cL_k$ is non-Hermitian, but since the Ornstein--Uhlenbeck process is reversible with respect to its stationary distribution $\rho_s$, the spectrum of $\cL_k$ is real and is conjugated to the spectrum of the following Hermitian operator \cite{touchette2017}:
\be
\cH_k=\frac{\sigma^2}{2}\p_{xx}-V_k(x)
\ee
which describes, up to a sign, the energy of a quantum particle in the potential
\be
V_k(x)= \frac{\gamma^2 x^2}{2\sigma^2}-\frac{\gamma}{2}-kx^\alpha.
\label{eqqpot1}
\ee
With the minus sign difference, $\lambda(k)$ therefore corresponds to the ground state energy (if it exists) of $\cH_k$ \cite{majumdar2002}. 

Other processes and observables can be analysed using the same method, working either with $\cL_k$ for general (possibly non-reversible) processes or with $\cH_k$ for reversible processes \cite{touchette2017}. If the potential $V_k(x)$ is not confining, then we formally expect $\z(\cL_k)=\infty$ for $k\neq 0$ and $I(a)=0$ by Legendre transform. In this case, the large deviation scaling of $P_\T(a)$ and $G(x,\T)$ is expected to be either not exponential (e.g., power-law in $\T$) or exponential but anomalous, as in \eqref{eqldtgen1}. The SCGF can also diverge for a confining potential because of boundary terms and the choice of initial distribution. This arises, for example, in the context of the so-called extended fluctuation relation when considering observables with a potential part that depends on the initial and final state, which still obey a normal LDP \cite{zon2003,sabhapandit2011,sabhapandit2012}. The anomalous large deviations described here are not related to this.

\subsection*{Low-noise approximation}

The probability distribution of $A_\T$ can be expressed in path integral form as
\be
P_\T(a) = \int dx_0\, \rho_s(x_0)\int d x_\T \int_{(0,x_0)}^{(\T,x_\T)} \Di[x]\, e^{-\SS_\T[x]}\delta (A_\T -a)
\ee
where $\rho_s$ is the stationary distribution of the Ornstein--Uhlenbeck process and $\SS_\T[x]$ is the classical action of that process shown in \eqref{eqaction} (see, e.g., \cite{nickelsen2011} and references therein). The delta function enforces the constraint that all trajectories contributing to $P_\T(a)$ must be such that $A_\T=a$ on the time interval $[0,\T]$.

In the low-noise limit, the path integral is dominated by the optimal path or instanton $\bx(t)$ that minimizes $\SS_\T[x]$ under the constraint $A_\T=a$. To obtain that path, we identify the Lagrangian density as
\be
L(x,\dx) = \frac{1}{2\sigma^2}(\dx + \gamma x)^2 - \beta x^\alpha,
\ee
which contains the Lagrange parameter $\beta$ that fixes the constraint. The associated Euler-Lagrange equations are found to be
\begin{align} 
\ddot x(t) &= \gamma^2x(t) - \beta\sigma^2\alpha x(t)^{\alpha-1}\nonumber\\
0 &= \dot x(0) - \gamma x(0)\nonumber\\
0 &= \dot x(\T) + \gamma x(\T),
\label{eqelesupp}
\end{align}
where the last two follow because of free boundary conditions imposed at $t=0$ and $t=\T$. 

These equations cannot be solved analytically. However, numerical solutions suggest that, for large $\T$, there is a unique instanton lying on the $H(x,p)=0$ manifold in phase space, where $H(x,p)$ is the Hamiltonian shown in \eqref{eqH1}, conjugated to $L$ with the momentum $p=(\dx + \gamma x)/\sigma^2$. Hamilton's equations read
\begin{align}
\dot x &= \frac{\p H}{\p p} = \sigma^2p - \gamma x\nonumber\\
\dot p &= -\frac{\p H}{\p x} = \gamma p - \alpha\beta x^{\alpha-1}.
\end{align}
Apart from the trivial (hyperbolic) fixed point $(0,0)$, another (stable) fixed point of this dynamics is
\be
\label{eqfp_supp}
(x^*,p^*) = \left(\left(\frac{\gamma^2}{\alpha\beta\sigma^2}\right)^{\frac{1}{\alpha-2}},\left(\frac{\gamma^\alpha}{\alpha\beta\sigma^{2\alpha-2}}\right)^{\frac{1}{\alpha-2}}\right).
\ee
For odd $\alpha>2$, these two fixed points are the only real fixed points, while for even $\alpha>2$ there is a third fixed point at $(-x^*,-p^*)$ due to the symmetry of the dynamics. We focus only on the positive fixed point $(x^*,p^*)$.

Since the instanton has zero energy, its action takes the form shown in \eqref{eqrateham}. The positive part of the $H=0$ manifold looping around the fixed point $(x^*,p^*)$ from the origin has two branches given by
\be
\bp_{1,2}(x) = \frac{\gamma x}{\sigma^2} \pm \frac{\gamma x}{\sigma^2}\sqrt{1-\frac{2\beta\sigma^2}{\gamma^2}x^{\alpha-2}},
\ee
and joined at the turning point $(\hx,p(\hx))$, shown in Fig.~\ref{figham}, where
\be 
\label{eqmaxx_supp}
\hx = \left(\frac{\gamma^2}{2\beta\sigma^2}\right)^\frac{1}{\alpha-2}.
\ee
The line integral in \eqref{eqrateham} is calculated separately on these two branches and yields
\be
\oint_{H=0} p(x) dx 	
= \frac{2\gamma\sqrt{\pi}}{\sigma^2(\alpha+2)}
\left(\frac{2\beta\sigma^2}{\gamma^2}\right)^{-\frac{2}{\alpha-2}}
\frac{\Gamma\big(\frac{2}{\alpha-2}\big)}{\Gamma\big(\frac{\alpha+2}{2\alpha-4}\big)}.
\ee
To determine $\beta$, we also evaluate the constraint along the two branches of the loop instanton:
\begin{align}
\T a &= \int_0^\T \bx(t)^\alpha dt\nonumber\\
&= \oint_{H=0} \frac{\bx^\alpha}{\dot\bx} d\bx\nonumber\\
&= \oint_{H=0} \frac{\bx^\alpha}{\sigma^2\bp(\bx)-\gamma\bx} d\bx\nonumber \\
&= \frac{2\sqrt{\pi}}{\gamma(\alpha-2)}
		\left(\frac{2\beta\sigma^2}{\gamma^2}\right)^{-\frac{\alpha}{\alpha-2}}
		\frac{\Gamma\big(\frac{\alpha}{\alpha-2}\big)}{\Gamma\big(\frac{3\alpha-2}{2\alpha-4}\big)},
\end{align}
which yields
\be
\beta(a) = (a\T)^{-\frac{\alpha-2}{\alpha}}\,\frac{\gamma^\frac{\alpha+2}{\alpha}}{2\sigma^2}
\left(\frac{2\sqrt{\pi}}{\alpha-2}\frac{\Gamma\big(\frac{\alpha}{\alpha-2}\big)}{\Gamma\big(\frac{3\alpha-2}{2\alpha-4}\big)}\right)^\frac{\alpha-2}{\alpha}.
\ee
Inserting these results back into the action \eqref{eqrateham}, we find as announced that $\SS_\T[x]\propto \T^{2/\alpha}$, so that $\xi=2/\alpha$, which leads, with \eqref{eqldtgen1}, to the low-noise rate function shown in \eqref{eqIscal}. 

As before, these results only give the positive part of $I(a)$, since this function is defined only for $a\geq 0$ when $\alpha$ is even, whereas $I(a)=I(-a)$ for $a\in\reals$ when $\alpha$ is odd. Moreover, the whole calculation applies only for $\alpha>2$; for $\alpha=1$ and $\alpha=2$, the low-noise calculation yields different instantons associated with normal large deviations.

\subsection{Exponential instanton approximation}

Numerical solutions of the Euler-Lagrange equations \eqref{eqelesupp} suggest that the instanton is well approximated by two exponentials with rate $\gamma$ about the middle time $\T/2$:
\be
\label{eqlargeTx}
\bx(t) \approx x_{\max} e^{-\gamma\left|\frac{\T}{2}-t\right|},
\ee
where $x_{\max}$ is the maximum position reached, fixed by the constraint $A_\T=a$, yielding
\be
x_{\max} =\left(a\T\frac{\alpha\gamma}{2}\right)^{\frac{1}{\alpha}}\left(1- e^{-\alpha\gamma\T/2}\right)^{-\frac{1}{\alpha}}.
\ee
Plugging this simple ansatz for the instanton into the action yields
\be
\SS_\T = (a\T\alpha)^\frac{2}{\alpha}\frac{\gamma^\frac{\alpha+2}{\alpha}}{4^{1/\alpha}\sigma^2} \frac{1-e^{-\gamma\T}}{\left(1- e^{-\alpha\gamma\T/2} \right)^\frac{2}{\alpha}} .
\ee
We see that this reproduces the correct $\T$ and $a$ scaling of the action, but not the prefactor \eqref{eqpref1} of the low-noise approximation of $I(a)$.

\subsection*{Monte Carlo simulations}

The numerical results presented in Fig.~\ref{figrtf1} were obtained using a direct Monte Carlo method by simulating $N$ samples (copies or replicas) of the Ornstein--Uhlenbeck process over the time interval $[0,\T]$ and by calculating $A_\T$ for each sample path. The process was simulated using a Euler-Maruyama discretization scheme with $\Delta t=0.01$ as the integration time-step. From the $N$ values of $A_\T$ obtained, we then constructed a normalized histogram $P_{\T,N}(a)$ of $A_\T$ and transformed that histogram to
\be
I_{\T,N}(a)=-\frac{1}{\T^\xi}\ln P_{\T,N}(a)
\label{eqldest1}
\ee
to get an estimate of the rate function \cite{touchette2011}. We repeated this procedure for different sample sizes $N$ and different integration times $\T$ to verify convergence. We also plot in Fig.~\ref{figrtf1} the rescaled rate function $\tI(a)=\sigma^2 I(a)$, since the low-noise result \eqref{eqIscal} is then independent of $\sigma$.

For $\alpha=3$, $\gamma=1$ and $\sigma\in \{0.5,0.75,1\}$, we found that $I_{\T,N}(a)$ becomes more or less constant for $\T\gtrsim 30$, after trying $\T=10,20,30$ and $40$ with $\xi=2/3$. For $\sigma=0.25$, convergence for the few points obtained was reached for $\T\gtrsim 20$. The sample size $N$ only determines the range of fluctuations over which $I(a)$ is obtained, and was set in simulations to $10^9$. Error bars were computed by constructing normal ``square-root'' error bars for the histogram and by transforming them according to \eqref{eqldest1}. They are shown for all data points in Fig.~\ref{figrtf1} but are, in most cases, smaller than the data points themselves.

The same simulations were used to produce the plot of Fig.~\ref{figinst1} by recording the trajectories of the process leading to a given fluctuation value \cite{grafke2015}, in this case $A_\T=0.45\pm 0.02$ for the parameters listed in the caption of that figure. The padding $\pm 0.02$ is added to make sure that paths are actually selected in simulations; it does not influence their shape in any significant way.

\subsection*{Rate function for even $\alpha$}

For even $\alpha>2$, $A_\T$ is a positive random variable whose mean $a^*$, as given by \eqref{eqmean1}, is strictly positive. In fact,
\be
a^*=\frac{\Gamma \left(\frac{\alpha +1}{2}\right) }{\sqrt{\pi}} \left(\frac{\sigma^2 }{\gamma}\right)^{\alpha /2}.
\label{eqmean2}
\ee

In this case, we find two regions of large deviations for $A_\T$. On the one hand, for $0\leq a<a^*$, $P_\T(a)$ has normal large deviations in $\T$, as in \eqref{eqldt1}, with a rate function $I(a)$ given by the Legendre transform of the dominant eigenvalue $\z(\cL_k)$, as described before. This arises because $V_k(x)$ then has a bound state for $k<0$, corresponding to values of $a$ below the mean $a^*$ for which $I'(a)=k<0$ \cite{touchette2009}. This normal region was studied for the Ornstein--Uhlenbeck process by Fatalov \cite{fatalov2006,fatalov2009}.

On the other hand, for $a>a^*$, $P_\T(a)$ has anomalous large deviations with $\xi=2/\alpha$, as found here, since $V_k(x)$ looses its bound states for $k>0$. The rate function is then approximated in the low-noise limit by our result \eqref{eqIscal} for $a>0$, since $a^*\ra 0$ in that limit, following the result \eqref{eqmean2} above.

For $a=a^*$, we simply have $I(a)=0$ under both scalings. Moreover, when $\alpha$ is odd, the normal region of large deviations disappears because $V_k(x)$ has no bound states for all $k\in\reals$.

\end{document}